\documentclass[aps,superscriptaddress,amsmath,amssymb,floatfix,showpacs,notitlepage,nofootinbib,preprintnumbers,twocolumn,prc]{revtex4-1}
\usepackage{graphicx,graphics,setspace,epsfig,color}
\usepackage[letterpaper, dvips,width=7.5in,height=8.5in,includemp=false]{geometry}
\usepackage[vcentermath]{}
\usepackage{epsf}
\usepackage{amscd}
\usepackage{amsmath}
\usepackage{graphicx}
\usepackage{dcolumn}
\usepackage{bm}
\usepackage{amssymb}
\usepackage{epsfig}
\usepackage{slashed}
\usepackage{lmodern} 
\usepackage[T1]{fontenc}

\newcommand{\mnras}{\textrm{Mon. Not. Roy. Astro. Soc.}}
\newcommand{\aap}{\textrm{Astron. \& Astrophys.}}

\newcommand{\beq}{\begin{equation}}
\newcommand{\eeq}{\end{equation}}
\newcommand{\bea}{\begin{eqnarray}}
\newcommand{\eea}{\end{eqnarray}}

\newcommand{\gcc} {{\rm g}/{\rm cm}^3}

\newcommand{\Ompl}{{\Omega_{\rm pl}}}
\newcommand{\kfe}{{k_{\rm Fe}}}
\newcommand{\ktf}{{k_{\rm TF}}}
\newcommand{\ktfsq}{{k^2_{\rm TF}}}
\newcommand{\vfe}{{v_{\rm Fe}}}
\newcommand{\efe}{{\epsilon_{\rm Fe}}}
\newcommand{\aem}{{\alpha_{\rm em}}}
\newcommand{\aemsq}{{\alpha^2_{\rm em}}}
\newcommand{\kfesq}{{k^2_{\rm Fe}}}

\def\aap{A\&A }
\begin{document}

\preprint{INT-PUB-16-003}

\title[title]{Thermal conductivity and impurity scattering in the accreting neutron star crust}
\author{Alessandro Roggero}
 \email{roggero@uw.edu}
\affiliation{Institute for Nuclear Theory, University of Washington, Seattle, WA}
\author{ Sanjay Reddy}
 \email{sareddy@uw.edu}
\affiliation{Institute for Nuclear Theory, University of Washington, Seattle, WA}
\affiliation{Department of Physics, University of Washington, Seattle, WA}

\begin{abstract}
We calculate the thermal conductivity of electrons for the strongly correlated multi--component ion plasma expected in the outer layers of neutron star's crust employing a Path Integral Monte Carlo (PIMC) approach. This allows us to isolate the low energy response of the ions and use it to calculate the electron scattering rate and the electron thermal conductivity. We find that the scattering rate is enhanced by a factor 2--4 compared to earlier calculations based on the simpler electron--impurity scattering formalism. This finding impacts the interpretation of thermal relaxation observed in transiently accreting neutron stars, and has implications for the composition and nuclear reactions in the crust that occur during accretion.  
\end{abstract}
\maketitle

\section{Introduction}
Observations of transient cooling in accreting neutrons stars and
magnetars \cite{Eichler:1989,Rutledge:2002} after outbursts has motivated
recent work to model the thermal evolution of the outermost regions of
the neutron star called the crust \cite{Shternin:2007,BrownCumming:2009,PageReddy:2012,PageReddy:2013}.  In these models the temporal structure of the 
x-ray light curves is set by the thermal conductivity of the crust. Here, the relevant density $ <10^{14}\,~\gcc$ 
and the expected temperature is in the range of $10^7-10^9$ K.  For densities $>10^6~\gcc$, electrons are relativistic and degenerate and they dominate transport phenomena. Under these conditions, nuclei are ionized and form a crystalline solid, and electron conduction is limited by electron-ion scattering. Since the typical electron wavelength is comparable to the distance between ions, interference between electron scattering off different ions is important.  Accounting for such interference under arbitrary ambient conditions in a multi-component plasma (MCP) is a challenging many-body problem because ions have large atomic number ($20\lesssim Z \lesssim 50$), and their dynamics is strongly correlated by Coulomb interactions at low temperature.  In this article we present the first quantum calculation of electron-ion scattering in MCPs and electronic transport properties. The results we present apply directly to the outer crust, however the technique we propose here and some insights also apply to matter at higher density in the inner crust where some neutrons drip out off nuclei to occupy states in the continuum \cite{Baym:1971ax}. 

In its simplest form, crustal matter is a one component plasma (OCP) of ions with charge $Z$ in the range $25-40$. The Coulomb interaction between ions is weakly screened by degenerate electron gas, and is given by $V(r)=Z^2\aem \exp{(-r/\lambda_e) }/r$ where  $\aem\simeq1/137$ is the fine structure constant, $\lambda_e = \sqrt{4 \alpha \vfe /\pi}~\kfe$ is the electron screening length, $\kfe$ and $\vfe$ are the electron  Fermi momentum and Fermi velocity, respectively.  We note that here and throughout this article we use natural units and set $\hbar= c=k_B=1$. The ground state of a one component plasma (OCP) is a BCC solid at low temperature, and electron scattering in this phase, including the effects of the single and multi-phonon processes, is well studied in terrestrial metals \cite{Ziman:1960}. Calculations of the electron thermal conductivity of an OCP over the full range of temperatures of interest to neutron star astrophysics are now available \cite{FlowersItoh,Baiko98,Potekhin99,OCPabbar}.  However, when  several species of ions are present these methods cannot be applied directly.  

In accreting neutron stars a diverse mix of nuclei is produced through rapid proton capture (rp-process) reactions at the surface \cite{Schtaz:2001}. We can expect several nuclear species to continue to coexist deeper in the crust because reaction pathways needed to process the rp-process ashes to the  ground state nucleus are blocked by large Coulomb barriers. Their evolution through electron capture reactions at shallow depths and pycno-nuclear fusion reactions deeper in the crust produces a complex multi-component mixture of ions with a wide range of $Z$  \cite{Gupta2007,Gupta:2008xw,Steiner:2012bq}. The description of electron scattering in such a multi-component plasma (MCP) at low temperature is the major goal of this study. To date electron scattering in MCP has only been studies in the classical limit where the De Broglie wavelength of the ions is small compared to the average inter-ion distance. In these earlier studies molecular dynamics was used to provide a quantitive description of the  thermal conductivity in the high temperature limit when $T \gtrsim \Omega_{\rm pl}$ \cite{ChuckMCP2009,Daligault2009} where 
\begin{equation}
\Ompl = \sqrt{\sum_k \frac{ 4 \pi \aem  Z_k^2  n_k }{M_k}}
\end{equation}
is the average plasma frequency. Here, for each nuclear species labelled by the subscript $k$,  $n_k$ is the number density, $M_k$ is the mass, $Z_k$ is the charge,  and the abundance $x_k=n_k/n_I$ where $n_I$ is the mean ion density.  

At lower temperatures of relevance to neutron stars, quantum effects cannot be neglected a priori. To include them we use the Path Integral Monte Carlo (PIMC) method (for a review see \cite{CeperlyRMP} ) to obtain the ion-ion correlation functions needed to calculate electron scattering rates and present first results of quantum calculations of the thermal conductivity.  For typical MCPs encountered in neutron stars, we find significant reduction of the thermal conductivity at low temperature compared to those obtained in earlier work based on treating the MCP as an OCP plus uncorrelated impurities  \cite{FlowersItoh,Baiko98,Potekhin99}. 

The article is organized as follows. In section \ref{sec:thermal_conductivity} we review well-known results for the electron thermal conductivity and its relation to the ion-ion correlation function. We discuss the quantum calculation of this correlation function in Euclidean time  using PIMC in section \ref{sec:PIMC} and present our results in section \ref{sec:results}. Finally we summarize and conclude in section \ref{sec:conclusions}.         

\section{Thermal conductivity}
\label{sec:thermal_conductivity}
In the crust where electrons are degenerate and weakly coupled, their thermal conductivity 
\begin{equation}
\kappa_e = \frac{1}{3}  C_V \vfe  \lambda_\kappa = \frac{\pi^2 T n_e}{3 \efe } \frac{1}{\nu_\kappa}
\end{equation}
where $C_V$ is the heat capacity  of the relativistic electron gas,  $\lambda_\kappa$ is the electron mean free path, and $\vfe=\kfe/\efe$ and $\efe =\sqrt{\kfesq+m_e^2}$ are the electron Fermi velocity and Fermi energy, respectively \cite{Ziman:1960,FlowersItoh}. The second equality is obtained by noting that the specific heat of a degenerate electron gas is $C_V = \pi^2 n_e T/(\vfe \kfe)$ where $n_e$ is the electron density and electron collision rate $\nu_\kappa=\vfe/\lambda_\kappa$. Under typical conditions electron-electron collisions are negligible and the total scattering rate  $\nu_\kappa=\nu_\kappa^{ee}+\nu_\kappa^{ei}\simeq \nu_\kappa^{ei}$ and in the following we will only consider the electron-ion scattering  process. This rate can be written as \cite{NandkumarPethick:1984}
\begin{equation}
\nu_\kappa = \nu^0_\kappa \frac{\langle Z^2 \rangle}{\langle Z \rangle}\Lambda_\kappa
\end{equation}
where
 \begin{equation}
\nu^0_\kappa = \frac{4 \aemsq  \efe}{3\pi}\,,
\end{equation}
is the characteristic collision frequency and 
\begin{equation}
\Lambda_\kappa=\int_0^{2 \kfe} dq ~h(q,\ktf,\kfe)~ S_\kappa(q)\,,
\label{eq:CoulombLog}
\end{equation}
is called the  Coulomb logarithm \footnote{Although in the plasma physics literature the Coulomb logarithm is usually defined without $S_\kappa(q)$ in the integrand, in the context of dense astrophysical plasmas, the quantity defined by Eq.~\ref{eq:CoulombLog} is also called the Coulomb logarithm.}. Here 
\begin{equation} 
h(q,\ktf,\kfe)= \frac{q^3}{\left( q^2 + \ktfsq\right)^2}  \left( 1 - \frac{q^2}{4\kfesq}\right) \,,
\end{equation}
and $\langle Z^n\rangle = \sum_i x_i Z^n_i$, $\kfe$ is the electron Fermi--momentum, $\ktf=1/\lambda_e$ is the Thomas--Fermi wave--vector and 
\begin{equation}
S_\kappa(q) = \int_{-\infty}^{\infty} d\omega \langle S'(\vec{q},\omega) \rangle_{\hat{q}} K(\beta \omega,q)
\end{equation}
is the structure factor for the thermal conductivity \cite{NandkumarPethick:1984} which contains all the information about ion--ion correlations.  $S'(\vec{q},\omega)$ is the dynamic structure factor with contributions from elastic Bragg scattering removed  because this contributes to the electron ground state wave-function and leads to their band structure but does not contribute to transport properties. $\langle \dots \rangle_{\hat{q}}$ denotes the average over the direction of unit vector $\hat{q} = \vec{q}/q$ and the  function  
\begin{equation}
K(z = \beta \omega,q) = \frac{z}{e^z-1} \left[ 1 + \frac{z^2}{\pi^2} \left( \frac{3p_F^2}{q^2} - \frac{1}{2}\right)\right].
\label{eq:Kfactor}
\end{equation}
incorporates the final state blocking of electrons and detailed balance that ensures typical energy transfers is of the order of the temperature. Finally, the second term in parenthesis incorporates energy exchanging small angle scattering contributions to the thermal conductivity.  

The charge--charge dynamic structure factor is the Fourier transform of the correlation function\begin{equation}
S(\vec{q},t) =\frac{1}{\langle Z^2 \rangle} \langle \rho^{\dagger}(\vec{q},t) \rho(\vec{q},t) \rangle_{\beta},
\end{equation}
where the factor $\langle Z^2 \rangle$ ensures the correct normalization $S(q,t=0)\xrightarrow{q\to\infty}1$ and $\langle \dots \rangle_{\beta}$ denote thermal averages at a temperature $1/\beta$. The charge density operator $\rho(\vec{q},t)$ is
\begin{equation}
\rho(\vec{q},t) = \frac{1}{\sqrt{N_i}} \sum_{i=1}^{N_i} Z_i e^{i\vec{q}\cdot\vec{r}_i(t)}
\end{equation}
with $Z_i$ and $\vec{r}_i(t)$ the charge and position of the i$^{th}$ ion at time $t$.

At high temperatures $T\gg\Ompl$ the bulk of the response is expected in the region where $z\ll1$. Here $K(\beta \omega,q)\approx 1$ and $S_\kappa(q)=S(q)$. However when $T<\Ompl$ this approximation fails and dynamical information is necessary to calculate $S_\kappa(q)$. With decreasing temperature, electron scattering only probes the response at small $|\omega|$ of order the temperature and it is imperative to identify the strength of  $S'(\vec{q},\omega)$ at $|\omega| \ll \Ompl$ to calculate the thermal conductivity.   

In the astrophysical context the MCP is often approximated as a perfect crystal with ions of charge $\langle Z \rangle$ at the lattice sites 
plus a randomly distributed impurity charge  $Z_j - \langle Z \rangle$  \cite{FlowersItoh,Itoh1993}. In this picture the total collision rate 
has two separate contributions $\nu_\kappa = \nu_\kappa^{ph} +
\nu_\kappa^{imp}$. The first contribution is due to the absorption or emission of phonons by electrons. These processes are  {\it inelastic} and consequently suppressed at low temperature. In contrast, impurity scattering is {\it elastic} and the  temperature independent scattering rate is given  by  
\begin{equation}
\nu_\kappa^{imp} = \nu^0_\kappa \frac{Q_{imp}}{\langle Z \rangle} \Lambda^{imp}_\kappa ,
\label{eq:impstdmodel}
\end{equation}
where 
\begin{equation} 
Q_{imp}= \langle Z^2 \rangle - \langle Z\rangle^2
\label{eq:Qimp} 
\end{equation} 
is called {\it impurity--parameter} and the Coulomb logarithm for uncorrelated impurities is 
\begin{eqnarray}
\Lambda^{imp}_\kappa&=& \int_0^{2p_F} dq~h(q,\ktf,\kfe)\,, \\
&=&\left( \frac{\aem}{\pi}+\frac{1}{2} \right)  \ln \left( \frac{\aem +\pi }{\aem} \right) -1\,. 
\label{eq:lambdaimpsimple}
\end{eqnarray}
In the neutron star context one finds that even for modest values 
of the $Q_{imp}\simeq 5$,  $\nu_\kappa^{imp} \gg  \nu_\kappa^{\langle Z \rangle}$ for typical temperatures in the range $10^6-10^8$ K in the denser regions of the crust where $\Ompl >T$ \cite{BrownCumming:2009}. 

The above mentioned approach to describe electron scattering in the MCP is only approximate. It neglects correlations between minority and majority species and correlations between minority species can also be important as nuclei with small $Z$ can cluster \cite{ChuckMCP2009}.  In the vicinity of an impurity we expect static distortions of the majority lattice. This would induce a non-periodic component to the Coulomb field which is bigger than the field associated with effective impurity charge $Z^{imp}-\langle Z\rangle$.  To account for these effects we require a method that can capture the dynamics of all species of ions on equal footing. In the classical limit, the dynamic  structure function $S'(\vec{q},\omega)$ of the MCP has been calculated using Molecular Dynamics (MD), while the static structure function $S(q)$ of MCP has been  calculated also with Classical Monte Carlo (CMC) methods  \cite{ChuckMCP2009,Daligault2009,OCPabbar}. In what follows we describe the PIMC technique needed to perform the quantum calculation of the response of a MCP.

 \section{Path Integral Monte Carlo simulations}
\label{sec:PIMC}
The Path Integral Monte Carlo method is an exact many--body technique to calculate equilibrium properties
of strongly interacting quantum particles with either Boltzmann or Bose statistics \cite{CeperlyRMP}. Because we expect ions to be 
spatially localized by strong Coulomb interactions even at low temperature when their thermal De Broglie wavelength are large, its 
a good approximation to treat them  as Boltzmann particles. There are two major advantages to using PIMC instead of CMC. First, it 
naturally incorporates zero-point motion of the ions and second, it is possible to obtain some dynamical information about the system by computing the euclidean (imaginary--time) correlation function
\begin{equation}
\begin{split}
F(\vec{q},\tau) &= \frac{Tr\left[\hat{\rho}^{\dagger} e^{-\tau\hat{H}} \hat{\rho} e^{-(\beta-\tau)\hat{H}}\right]}{Tr[e^{-\beta\hat{H}}]}\\
&= \int_{-\infty}^{\infty} d\omega e^{-\tau\omega} S(\vec{q},\omega)\\
\end{split}
\label{eq:Fqtau}
\end{equation}
where $\tau\in[0,\beta]$ is the imaginary--time interval. The latter will turn out to be particularly useful in describing $S(\vec{q},\omega)$ at $\omega \approx 0$. As discussed earlier the elastic response with $\omega \approx 0$  provides the temperature independent contribution to electron scattering and dominates at low temperature. 

The traces in Eq.~\eqref{eq:Fqtau} are performed over the $3N_{i}$--dimensional configuration 
space of the ions and the full path $[0,\beta]$ is split into $M$ slices. This procedure allows reliable approximations for the density matrices 
$\langle R \lvert \exp(-\Delta\tau\hat{H}) \rvert R' \rangle$ at the higher (inverse) temperature $\Delta\tau=\beta/M$. For the conditions explored in this study, we found the
primitive approximation very accurate by comparing it to the exact two--particle density matrix obtained in the Feynman--Kac approach (see \cite{CeperlyRMP} for details).
The final result of the PIMC calculation is $F(\vec{q},\tau)$ at $M$ discrete values of imaginary--time $\tau$ and for a large number of momentum 
transfers compatible with the periodic boundary conditions of the simulation box, ie. $\vec{q}=(2\pi/L)(n_x,n_y,n_z)$ with $L=(N_{i}/n_{i})^{1/3}$ 
the length of the box and $n_x,n_y,n_z$ integers. 

The only systematic bias present in the current computation is due to finite--size effects which in this case are mostly caused by the long--range nature 
of the interaction. However, due to screening a detailed resolution of long distance effects through Ewald summations \cite{Ewald} is unnecessary. We find that summing over image charges in two nearest neighbor cells in all directions is adequate (this corresponds to including a total $26$ cells surrounding the simulation box). We have checked the converge of this procedure for both energies and structure factors using systems composed of different number of ions $N_{i}$ ranging from $686$ to $2662$ and found convergence when $N_{i} \gtrsim 1024-1458$ in good agreement with earlier findings in Ref.~\cite{OCPabbar}. In what follows we present results obtained with systems with $N_{i} = 2000$.

\section{Multi-component Plasmas in Accreting Neutron Stars}
In accreting neutron stars rapid proton capture reactions can produce a diverse mix of nuclei at the surface with $Q_{imp}\simeq 100$  \cite{Schatz:1999kx,Schtaz:2001}. As this mixture is incorporating into the outer
crust phase separation and electron capture reactions purify the mix somewhat and the impurity parameter in the outer crust is 
expected to be in the range $Q_{imp}\approx 10$ (cf. \cite{Gupta2007,Chuck2007}). In this study we approximate the complicated 
composition found in \cite{Gupta2007} with a one component plasma with a charge $Z=\langle Z\rangle$ and two- and three--component 
system labelled MCP1, MCP2 in Tab.~\ref{tab:zax} to mimic $Q_{imp}\approx 15$ and  $Q_{imp}\approx 30$. In addition we considered 
a two--component system, labelled MCP3, with $Q_{imp}\lesssim 4$ motivated by findings in Ref.~\cite{BrownCumming:2009,Turlione2015}. The ion charges, their abundances denoted by $x$ and the impurity parameters are shown in Tab.~\ref{tab:zax}. In all cases the mass density has been fixed to $\rho=10^{10}\,~\gcc$.
\begin{table}[h]
\centering
\begin{tabular}{lc|ccc}
& & Z & A  & x\\\hline
OCP& & $32.74$ & $97.71$ & $1$\\\hline
MCP1:&$Q_{imp}=15.34$ & $32.74$ & $97.71$ & $0.76$\\
&& $23.57$ & $70.35$ & $0.24$\\\hline
MCP2:&$Q_{imp}=33.75$ & $32.67$ & $97.73$ & $0.6898$\\
&& $23.85$ & $71.33$ & $0.2685$\\
&&$8.556$&$25.33$&$0.0417$\\\hline
MCP3:&$Q_{imp}=2.16$ & $32.74$ & $97.71$ & $0.86$\\
&& $28.5$ & $70.0$ & $0.14$\\
\end{tabular}
\caption{Parameters for the systems considered in this work: charge $Z$, mass number $A$ and fraction $x=n_x/n$. 
For the MCP system we list also the $Q_{imp}$ parameter.}
\label{tab:zax}
\end{table}

As can be expected at low temperature we find a large number of metastable states with large barriers and setting up the initial conditions is quite challenging.  
A careful analysis of the equilibrium configuration in a low temperature MCP is beyond the scope of this work, and would require 
an annealing algorithm to evolve to the low temperatures structure starting from a high temperature liquid configuration. Such 
calculations have been performed in the past using molecular dynamics simulations \cite{Chuck2007,ChuckMCP2009,ChuckMCP2009_second} 
where it was found that the ground state was a BCC crystal composed by ions with large charge at regular lattice sites and low--Z ions
where found to occupy interstitial regions. This is not surprising since the Coulomb interaction is quite soft and allows for the diffusion in the solid phase that ensures amorphous structures to relax to crystalline state \cite{Hughto2011}.

Motivated by these findings we initialize the ions on lattice--sites of a perfect BCC crystal which is then distorted by applying random displacement 
to all the particles. The system is allowed to relax to its ground--state after which statistics for observable start to be taken. In a multi--component
system we also choose the type of ion on a given lattice at random, and during equilibration ions are allowed to interchange location with others of different species.
As can be seen from the presence of BCC Bragg peaks in the structure factors in Fig.~\ref{fig:sofqs}, the equilibrium configuration attained with this procedure is always an ordered BCC crystal phase. When ions with very small Z are present as in the MCP2, we found that in our simulations they remained as substitutional impurities on lattice sites. This may be an artifact of our simple initialization procedure and it would be interesting
to properly explore annealing in future using replica--exchange methods \cite{Marinari1992,Sugita1999}. However, since this only affects the distribution of ions with very small $Z$ species we do not expect  their dynamics to dominate the charge-charge correlation function. 

\section{Results}
\label{sec:results}
We first present results for the static structure factor $S(q)$ of the OCP and the MCPs for the compositions shown in Tab.~\ref{tab:zax} at three different temperatures $T=0.02,0.002,0.0004 $ MeV corresponding to $T/\Ompl\approx0.9,0.09,0.02$. Calculations employing Classical Monte Carlo (where the path--integral is restricted to a single time--slice) were also performed to asses the importance of zero--point motion. For the OCP we confirm earlier results in \cite{OCPabbar}  where it was found that the static structure function $S(q)$ is underestimated in the classical case when $\eta=T/\Ompl < 1$. 

\begin{figure}[h]
\begin{center}
\includegraphics[scale=0.35]{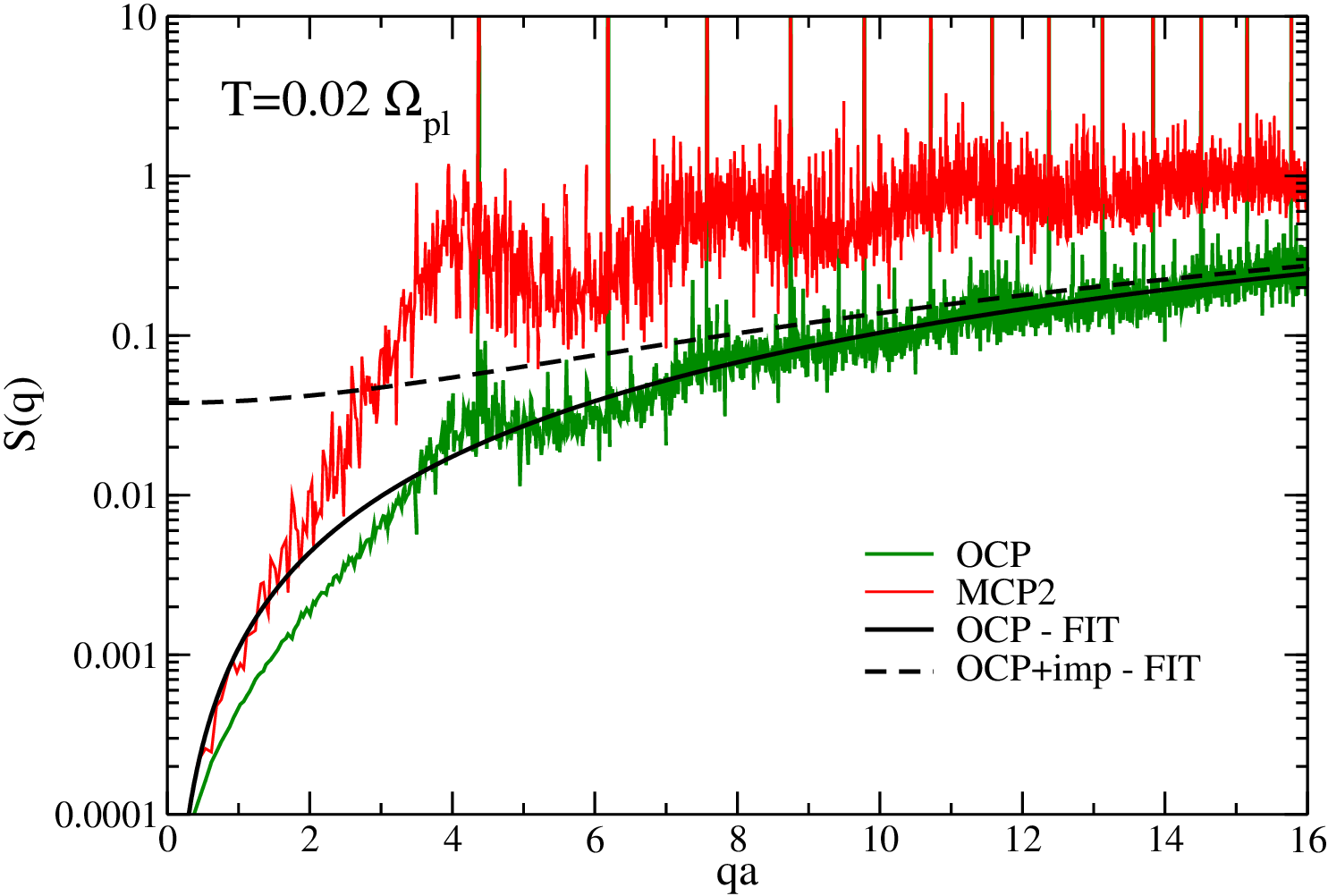}
\end{center}
\caption{Static charge--charge structure factor at $T=0.02~ \Ompl$ for the OCP and MCP2 systems (cf. Tab~\ref{tab:zax}). The $\approx1\%$ 
statistical error is not shown for clarity. \label{fig:sofqs}}
\end{figure}

In Fig.~\ref{fig:sofqs} we show  $S(q)$ for the OCP and MCP2 systems at the lowest temperature $T=0.02~ \Ompl$. It is apparent from the appearance of Bragg peaks that the underlying BCC structure persists in this MCP in agreement with earlier results obtained using MD\cite{ChuckMCP2009_second}.
For the OCP it is well known that the one phonon approximation provides a good description of S(q) at low temperature and that multi-phonon effects can be accounted for within the Harmonic Approximation. In \cite{Baiko98}, a simple analytic formula was developed based on the Harmonic Approximation (HA) to describe the static structure factor of an OCP and is shown as the solid black curve labelled OCP-FIT in the figure. Explicitly this is given by 
\begin{equation}
S_{\rm OCP-HA}'(q) = 1-e^{-2W(q)}
\label{eq:sofq_ha}
\end{equation}
where the Debye--Weller factor was found to be of the following simple analytic form 
\begin{equation}
W(q) = \frac{2a^2q^2}{3 \Gamma  }\left(u_{-2} + \frac{u_{-1}}{2\eta}~ e^{-9.1\eta}\right)\,, 
\end{equation}
to describe systems over a wide range of temperature \cite{Baiko95}.  Here $\eta=T/\Ompl$ and $u_n=\langle (\omega^n_{\nu}/\Ompl)\rangle_{ph}$ are the moments of the phonon spectrum. For a BCC lattice $u_{-1}=2.8$ and $u_{-2}=13.0$  \cite{Pollock73}.  For the MCP2 the simple impurity model predicts 
\begin{equation}
S^{\rm OCP + imp} (q)=\frac{Q_{imp}}{\langle Z^2 \rangle} + \frac{\langle Z \rangle^2}{\langle Z^2 \rangle}S_{\rm OCP-HA}'(q) 
\end{equation} 
since the contribution due to uncorrelated random impurities is additive. This is shown as the black-dashed curve in the figure and is labelled (OCP+imp-FIT). 

Apart from sharp features coming from the Bragg peaks, the analytical fit is on average in good agreement with our numerical result for $qa \gtrsim 3$. However, the comparison between our results for the MCP2 and those obtained using the OCP + impurity fit show important differences. Impurity effects included through Eq.~\ref{eq:impstdmodel} significantly underestimates the contribution to $S(q)$ and this is in qualitative agreement with results found using MD simulations at higher temperatures \cite{ChuckMCP2009,ChuckMCP2009_second}. 

\begin{figure}[h]
\begin{center}
\includegraphics[scale=0.35]{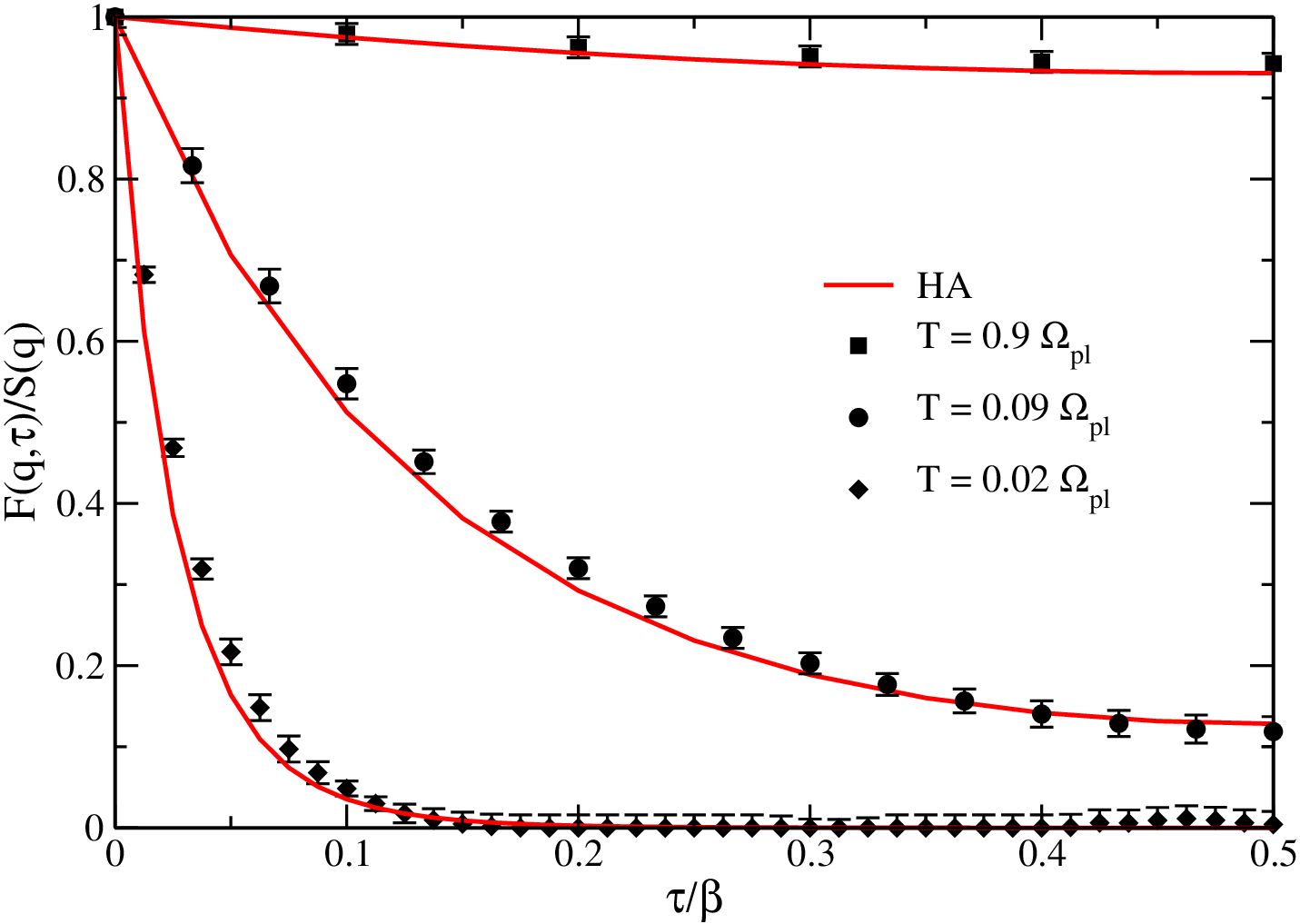}
\end{center}
\caption{Normalized euclidean charge--charge correlation functions for the OCP at different temperature for momentum transfer $qa\approx3.5$: black points are PIMC data, red curves are the HA results. }
\label{fig:eucOCP}
\end{figure}
We now turn to discuss how we obtain informations about the dynamical structure factor $S(q,\omega)$ from the Euclidean response function 
$F(q,\tau)$.  From Eq.~\eqref{eq:Fqtau} one can in principle try to perform a numerical Laplace transform inversion starting 
with the PIMC data to obtain the frequency dependent response. However, as is well known \cite{McWhirter,Talbot}, this numerical inversion 
is an ill-posed problem in the sense that arbitrarily small perturbations in the initial data can give rise to arbitrarily large deviations 
in the final answer. We attempted this inversion with different techniques and regularization schemes (\cite{Magierski2012,Roggero13} and 
references therein) with little success. This is because $S_\kappa(q)$ is very sensitive to response in the low energy at low temperatures, 
whereas the response obtained from numerical inversion is expected to be accurate in the region where the spectrum has maximum strength, and these two regions do not overlap significantly for most momentum transfers.

Due to the above mentioned problems, we choose instead to constrain models for the  dynamical structure factor $S(q,\omega)$ with 
PIMC data for $F(q,\tau)$. First, for the OCP we calculate the Euclidean response function using $S(q,\omega)$ obtained in the 
HA and taking the Laplace transform as defined in Eq.~\eqref{eq:Fqtau}. In Fig.~\ref{fig:eucOCP} we compare the (normalized) 
euclidean response for charge--charge fluctuations obtained in this way with the predictions of PIMC for a representative fixed 
momentum transfer $qa\approx3.5$ at three different temperatures for the OCP system. The excellent agreement between PIMC results 
and HA model is expected and confirms that the harmonic approximation works quantitatively. The striking feature is the rapid decrease in 
 $F(q,\tau)$ for large $\tau \simeq \beta/2$ at low temperature.  This simply reflects the fact that in the OCP the low energy response 
 due to phonon excitations is highly suppressed at low temperature and demonstrates that PIMC data for $F(q,\tau)$ at $\tau \simeq \beta/2$ provides powerful means to extract the response at very low energy when $T \ll \Ompl$.  
 
\begin{figure}[th]
\begin{center}
\includegraphics[scale=0.35]{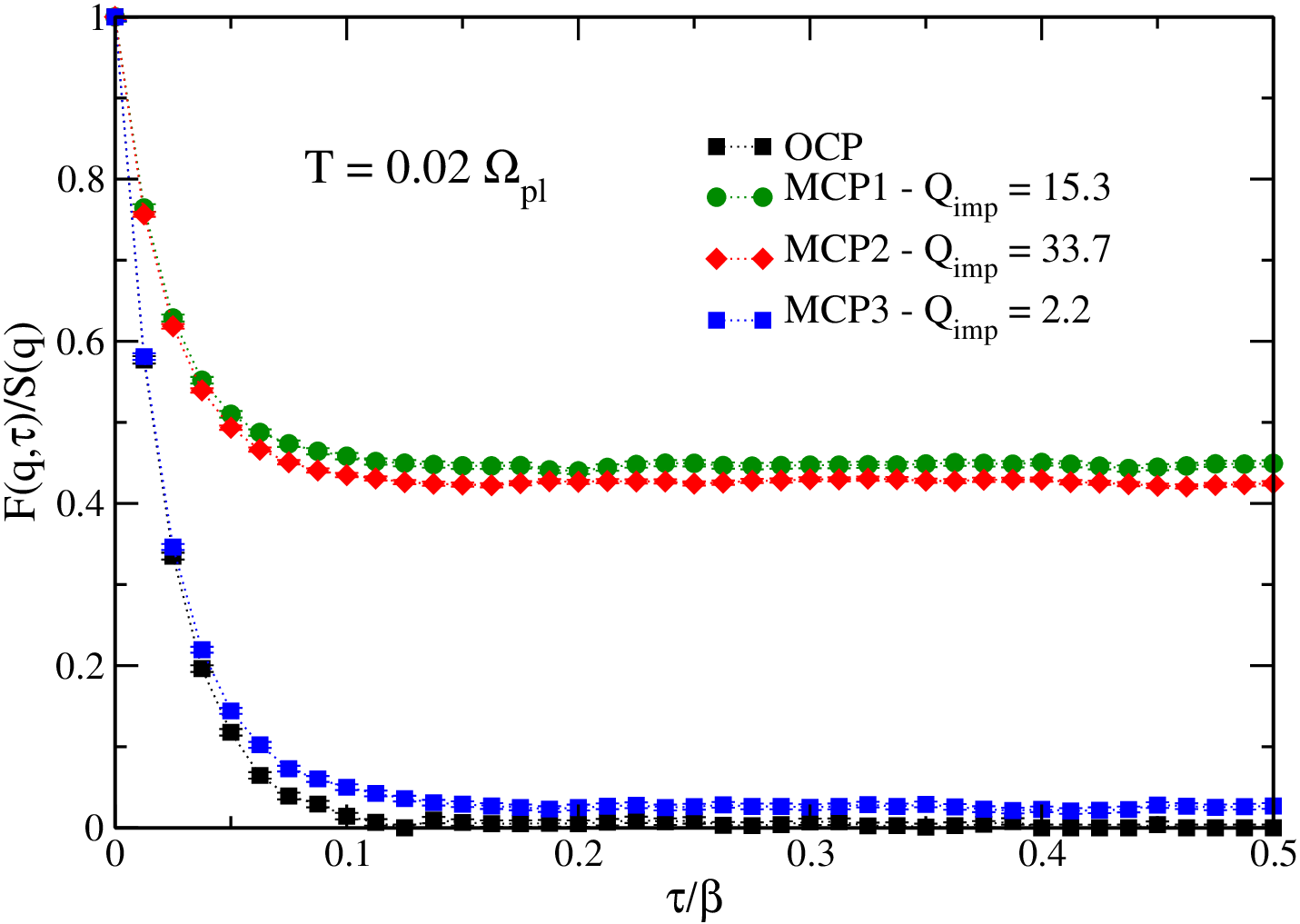}
\end{center}
\caption{Normalized euclidean charge--charge correlation functions for the OCP and three different mixtures MCP1, MCP2 and MCP3 (cf. Table~\ref{tab:zax}) at a 
temperature $T = 0.0004 MeV \approx 0.02 \Ompl$. The momentum transfer is $qa\approx2.2$.\label{fig:eucMCPvsOCP}}
\end{figure} 
 
The Euclidean correlation function  $F(q,\tau)$ in the MCP is shown in Fig.~\ref{fig:eucMCPvsOCP} for a momentum transfer $qa\simeq 2.2$ at $T=0.02 \Ompl$. A 
comparison between the results for the OCP (shown as black squares) and the MCP at $\tau \simeq \beta/2$ shows as expected that imperfections in the lattice produce 
a significant response at low energy.  The presence of excitations at $\omega\approx0$ can be directly inferred by the magnitude of the correlator at the largest 
imaginary--time separation $\tau=\beta/2$: in the OCP, $F(q,\beta/2)$ goes to zero at large imaginary--time indicating the absence of low energy excitations 
for $\omega\ll \beta^{-1}$, whereas in the MCP systems the persistence of correlations indicate cleanly the presence of significant strength at these low energies. 
This low energy response can be thought of as compositional modes with long relaxation times, corresponding to nearly static distortions of the periodic potential 
expected in the OCP.  Molecular dynamics studies by Caballero et al. \cite{Caballero2006} also found additional strength at very low energy in MCP. 

The absence of structure in  $F(q,\tau)$ for $\tau \gtrsim 0.1 \beta$ seen in Fig.~\ref{fig:eucMCPvsOCP} indicates that there is a clear separation between the low 
and high energy responses. This motivates us to write the dynamical structure factor at low temperature as 
\begin{equation}
S(q,\omega) = S_{ph}(q,\omega) + S_{imp}(q) \delta(\omega) .
\label{eq:sqwmcp}
\end{equation}
where we separate the contributions at finite frequency due to excitation of phonons and an elastic contribution due to nearly static lattice imperfections generated 
by the presence of impurities. Taking its Laplace transform we obtain 
 \begin{equation}
S_{imp}(q)=F(q,\tau)  - F_{ph}(q,\tau) \,. 
\label{eq:fqwmcp}
\end{equation}
In general the extraction of $S_{imp}(q)$ will rely on a model for $F_{ph}(q,\tau)$ that can capture the $\tau$ dependence of the full Euclidean correlator, however 
since $F_{ph}(q,\tau) \rightarrow 0$ in the low temperature limit 
\begin{equation}
 S_{imp}(q)= \lim_{\beta \to \infty} F(q,\beta/2)\,.  
\label{eq:simpofq}
\end{equation}
This identification is one important result of our study and provides a simple and robust strategy to calculate the thermal conductivity of complex mixtures encountered 
in accreting neutron stars for $T \ll \Ompl$. In practice we find that for $T < 0.1\Ompl$ the contribution of the phonons is small 
and there is no difference between using Eq.~\eqref{eq:simpofq} and Eq.~\eqref{eq:fqwmcp} to extract $S_{imp}(q)$ independently of the model for $F_{ph}(q,\tau)$. 
We tried calculating $F_{ph}(q,\tau)$ using $S(q,\omega)$ for the average $Z$ OCP and for the linear mixing model \cite{Daligault2009,Potekhin99} for the MCP 
in the harmonic approximation and found negligible differences. In the following we will denote results obtained using Eq.~\eqref{eq:fqwmcp} or Eq.~\eqref{eq:simpofq} as "EUC".

For high temperature when $T\gtrsim \Ompl$ we have already noted that $S_\kappa(q)=S(q)$ and dynamical information contained in $S(q,\omega)$ is not necessary to calculate the thermal conductivity.  At moderate 
temperature $0.1\Ompl<T\leq \Ompl$ the situation is complicated because the elastic contribution from phonons can be significant for large $\tau$ and the extraction of the contribution due to impurities will have some 
model dependence. Calculations of $S(q,\omega)$ in MCP using molecular dynamics in Ref.~\cite{Caballero2006} showed that the finite frequency contribution to the dynamical structure factor is very similar to the $S_{ph}(q,\omega)$ expected in the OCP with $Z=\langle Z\rangle$. This allows us to approximate the impurity contribution at moderate temperatures as 
\begin{equation} 
S_{imp}(q)=S_{MCP}(q)-S_{OCP-HA}(q)\,
\label{eq:staticimp}
\end{equation} 
where $S_{MCP}(q)$ is calculated using PIMC. We note that Eq.~\eqref{eq:staticimp} is remarkably accurate as its predictions at low temperature are consistent with the model--independent extraction obtained using Eq.~\eqref{eq:simpofq}.

\begin{figure}
\begin{center}
\includegraphics[scale=0.35]{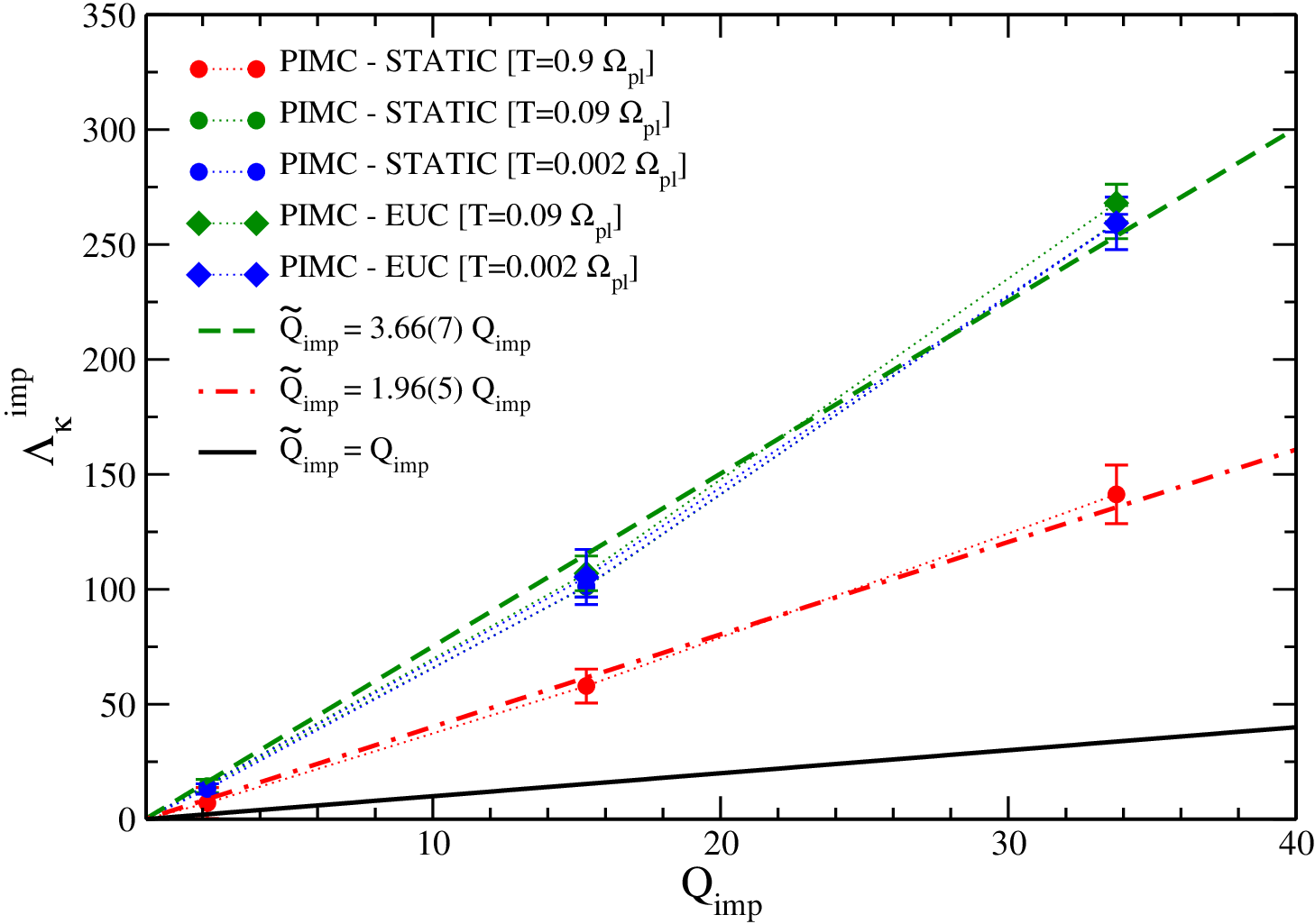}
\end{center}
\caption{Coulomb log for impurity scattering calculated using $S_{imp}(q)$  from the relation in Eq.~\eqref{eq:simpofq} and Eq.~\eqref{eq:staticimp} are shown and labelled PIMC-EUC and PIMC-STATIC, respectively. The simple impurity model prediction from Eq.~\eqref{eq:impstdmodel} is also shown (black solid line). \label{figQeff1}}
\end{figure}

\begin{figure}[h]
\begin{center}
\includegraphics[scale=0.35]{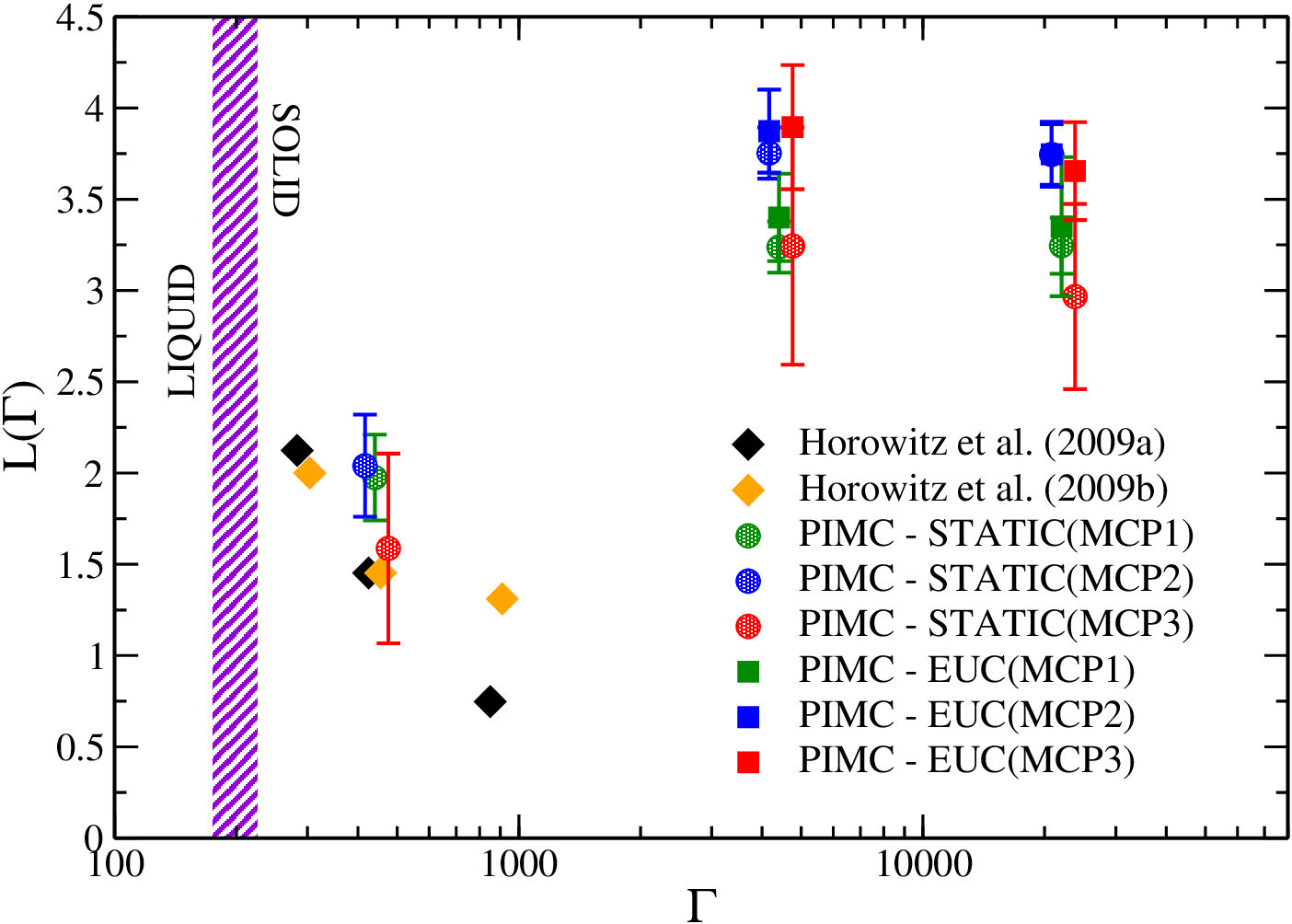}
\end{center}
\caption{ Impurity parameter correction factor $L(\Gamma)$ (cf. Eq.~\eqref{eq:Lofeta}) as a function of $\Gamma$ for the 3 mixtures considered in this work.   Points labelled PIMC-EUC and PIMC-STATIC are obtained by calculating the Coulomb logarithm using $S_{imp}(q)$  from the relation in Eq.~\eqref{eq:simpofq} and Eq.~\eqref{eq:staticimp}, respectively. Also shown the MD results from \cite{ChuckMCP2009} 
(black diamonds) and \cite{ChuckMCP2009_second} (orange diamonds), together with the liquid--solid transition region as a purple band. 
\label{figQeffm}}
\end{figure}

In Fig.~\ref{figQeff1} we plot the Coulomb logarithm  associated with the impurity contribution and defined by 
\begin{equation}
\begin{split}
\Lambda^{imp}_{\kappa} &= \int_0^{2p_F} dq~ h(q,k_{TF},p_F) S_{imp}(q)
\end{split}
\label{eq:impfracfromEta}
\end{equation}
for various temperatures and for three multi-component mixtures presented in Tab.~\ref{tab:zax}. 
It is interesting to note that  $\Lambda^{imp}_{\kappa}$ is approximately linear in $Q_{imp}$ and this allows us to define an effective impurity parameter 
\begin{equation}
 \widetilde{Q}_{imp} =L Q_{imp}\,.
\label{eq:Lofeta}
\end{equation}
The value extracted for $L$ is found to be temperature dependent: we obtain $L=1.96(5)$
for $T\approx0.9\Ompl$ and asymptotes to $L=3.66(7)$ in the low temperature limit. To explore the temperature dependence we plot $L$ as a function of the Coulomb coupling parameter $\Gamma= \langle Z \rangle^2/a T$ in Fig.~\ref{figQeffm}. The figure also shows results obtained from molecular dynamics simulations of mixtures with 
$Q_{imp}=38.9$ \cite{ChuckMCP2009} (black diamonds in the figure) and $Q_{imp}=22.54$ \cite{ChuckMCP2009_second} (orange diamonds) performed at a much higher density $\rho=10^{13}$ g/cm$^3$. It is important
to stress that, despite the higher density, these calculations where targeted to describe the outer crust, and therefore neglect effects of finite size of the ions as well as the presence of interstitial neutrons.
There is good agreement between our results and those obtained with molecular dynamics at smaller value of $\Gamma \simeq 300$ but the trend with increasing $\Gamma$ is different. The origin of this difference is unclear and warrants further work. 

\section{Summary and Conclusions}
\label{sec:conclusions}
We have calculated the Euclidean charge-charge correlation  of MCP using PIMC simulations and used it to extract the low energy response needed to calculate the electron thermal conductivity of high density matter at low temperatures encountered in neutron stars. The behavior of the Euclidean correlation function at large imaginary time was found to be dominated by slowly varying compositional modes associated with impurity induced lattice distortions. These contribute to the low energy strength in the ion dynamic structure and dominates electron scattering when $T \le \Ompl$. The role of impurity scattering in MCP has been studied in previous work in the high temperature classical regime using MD simulations  \cite{Caballero2006}. Our study is the first to extend the results to lower temperatures of relevance to neutron stars and includes quantum effects. The results we obtain are in agreement with the MD results at higher temperature which found that the simple impurity scattering formula underestimates the low energy response   
\cite{ChuckMCP2009,ChuckMCP2009_second}. However, we find that at low temperature this enhancement is substantially larger and implies that when $T\ll \Ompl$ the electron collision rate in typical MCP encountered in accreting neutron stars is about  factor of 4 larger than earlier estimates based on the simple impurity scattering formula. 

We have performed calculations for three MCP characterized by the impurity parameter $Q_{imp}\simeq 2, 15$ and $30$ at various temperatures and find that the electron scattering rate 
increases linearly with $Q_{imp}$. This allowed us to define a temperature dependent effective impurity parameter $\widetilde{Q}_{imp} =L(\Gamma) Q_{imp}$ which can be used to calculate 
the electron thermal conductivity using the simple impurity formula in Eq.~\eqref{eq:impstdmodel}. In our simulations we found that the ions are highly localized and the Bragg peaks 
associated with a BCC lattice persists. This suggests that electron scattering is predominantly due to distortions of the lattice rather than due to scattering of randomly distributed 
impurity charge and is a likely explanation for the enhanced response at low energies that we observe. When some fraction of the impurities occupy interstitial spaces through thermal motion these distortions can be screened 
and offer an explanation for the reduction in the $L$ at higher temperature.      

Finally, we comment on the implications of our study for the interpretation of observed thermal relaxation in accreting neutron stars. Models that best fit observations require a relatively large thermal conductivity in the inner crust with $\widetilde{Q}_{imp}\lesssim 4$  \cite{Shternin:2007,BrownCumming:2009,PageReddy:2013,Turlione2015}.  Although our results do not directly apply to the inner crust because of the presence of dripped neutrons, it is reasonable to expect that qualitative trends will be similar since the Coulomb interaction will continue to be the dominant forces between ions. The enhanced scattering rates implied by $L>1$ indicates that the impurity parameter $Q_{imp} $ in the inner crust can be a factor of a few smaller than $\widetilde{Q}_{imp}$. With future observations and improved modeling it may be possible to obtain useful constraints on the low temperature conductivity of the outer crust. Our calculations combined with such input will provide useful constraints on the  composition of the outer crust of an accreting neutron star. 

\begin{acknowledgments}
We would like to thank S.\ Abbar, J.\ Carlson, H.\ Duan and F.\ Pederiva for useful discussions. 
The work of S. R. was supported by the DOE Grant No. DE-FG02-00ER41132 and by the Joint Institute for Nuclear Astrophysics (JINA-CEE). The work of A.\ R. was supported by NSF Grant No.\ AST-1333607. 
Most of the intensive computations have been performed at NERSC thank to a Startup allocation.
\end{acknowledgments}

\bibliography{paper}
\end{document}